\documentclass{article}
\usepackage[utf8]{inputenc}
\usepackage[noadjust]{cite}
\usepackage{graphicx}
\usepackage{geometry}
\usepackage[noblocks]{authblk}
\usepackage{setspace}
\usepackage{color,xcolor}
\usepackage{ulem}
\usepackage{hyperref}

\begin{document}

\title{Coherent optical communications using coherence-cloned Kerr soliton microcombs}
\small
\author[1,3]{Yong Geng}
\author[1,3]{Heng Zhou \thanks{zhouheng@uestc.edu.cn}}
\author[1]{Wenwen Cui}
\author[1]{Xinjie Han}
\author[1]{Qiang Zhang}
\author[1]{Boyuan Liu}
\author[2]{Guangwei Deng}
\author[2]{Qiang Zhou}
\author[1]{Kun Qiu}
\affil[1]{Key Lab of Optical Fiber Sensing and Communication Networks, University of Electronic Science and Technology of China, Chengdu 611731, China}
\affil[2]{Institute of Fundamental and Frontier Sciences, University of Electronic Science and Technology of China, Chengdu 611731, China}
\affil[3]{These authors contributed equally: Yong Geng, Heng Zhou}
\date{}
\maketitle

\begin{abstract}
\begin{spacing}{1.1}
\large
Dissipative Kerr soliton microcomb has been recognized as a promising on-chip multi-wavelength laser source for fiber optical communications, as its comb lines possess frequency and phase stability far beyond independent lasers. In the scenarios of coherent optical transmission and interconnect, a highly beneficial but rarely explored target is to re-generate a Kerr soliton microcomb at the receiver side as local oscillators that conserve the frequency and phase property of the incoming data carriers, so that to enable coherent detection with minimized optical and electrical compensations. Here, by using the techniques of pump laser conveying and two-point locking, we implement re-generation of a Kerr soliton microcomb that faithfully clones the frequency and phase coherence of another microcomb sent from 50 km away. Moreover, leveraging the coherence-cloned soliton microcombs as carriers and local oscillators, we demonstrate terabit coherent data interconnect, wherein traditional digital processes for frequency offset estimation is totally dispensed with, and carrier phase estimation is substantially simplified via slowed-down phase estimation rate per channel and joint phase estimation among multiple channels. Our work reveals that, in addition to providing a multitude of laser tones, regulating the frequency and phase of Kerr soliton microcombs among transmitters and receivers can significantly improve coherent communication in terms of performance, power consumption, and simplicity. 
\newline
\end{spacing}
\end{abstract}

\begin{spacing}{1.6}
\section*{Introduction}
\large
Wavelength divided multiplexing (WDM) optical coherent transmission greatly enhances the capacity and spectral efficiency of fiber communication by modulating information onto both the amplitudes and phases of a multitude of laser carriers at the transmitter, and demodulating the encoded information at the receiver through coherently mixing the data signals with matched local oscillators (LO) \cite{Ezra}. Frequency and phase coherence between the carrier and LO lasers thus play a crucial role in determining the performance of coherent data receiving. To date, most commercial systems still use independent carrier and LO lasers, which have weak mutual coherence (Fig. 1a left) and entail large guard band and power-hungry digital signal processing (DSP) to gauge their frequency and phase uncertainties \cite{Xiang}. Optical frequency comb consisting of a large quantity of even spaced and phase locked laser tones can provide spectral stability orders of magnitude higher than individual lasers (Fig. 1a right) \cite{Diddams}, thus being considered as a promising laser source for coherent WDM transmission and interconnect. Their strengths to carry massive parallel data channels have already been demonstrated in various of optical frequency comb platforms, including electro-optical (EO) modulating comb \cite{mazur,Lars,Temprana}, nonlinear broadened comb \cite{Vahid,Hao}, mode-locked fiber laser comb \cite{Hillerkuss}, semiconductor gain-switched laser comb \cite{Joerg}, and dissipative Kerr soliton (DKS) microcomb \cite{Herr1,Pablo,Mikael,Attila,Bill,BaoCJ}. Therein, DKS microcomb generated in nonlinear optical microcavity has evoked special interests thanks to its unique features including large frequency spacing \cite{Herr1,Pablo}, ultra-broadband spectrum \cite{Brasch}, high stability \cite{Junqiu}, excellent SWaP (size, weight and power) factors and compatibility for chip integration \cite{Brian,Arslan,Boqiang}. It had been reported that chip-scale DKS microcombs can simultaneously provide more than 100 laser tones to transmit coherent data signals with line rate up to 55 Tbit s$^{-1}$ \cite{Pablo}. 

On the other hand, to employ DKS microcombs in coherent communication networks, it is of great importance to re-generate a LO microcomb at the receiver side that inherits the frequency and phase coherence of the transmitted data carrier comb \cite{Juned}, which, however, has not been comprehensively investigated to date \cite{Jae,Peicheng}. In fact, the generation dynamics and physical characteristics of DKS microcomb make it an ideal platform to realize coherence-cloned comb re-generation among transmitters and receivers \cite{Herr1,Brasch,Jae,Peicheng}. First, DKS microcomb is commonly generated by a single continuous-wave pump laser, which directly set the central frequency $f_c$ of the whole comb spectrum \cite{Herr1}. Second, the mode spacing $f_{spc}$ of a DKS microcomb is preset by the cavity geometry and can be finely adjusted by tuning the pump-cavity frequency detuning (and fundamentally the nonlinear phase-matching) \cite{Jae, Chengying}. Third, in soliton mode-locked state, the phases of all the DKS microcomb lines $\phi_m (m=\pm1, 2, 3,...)$ uniformly align to the phase of the pump laser $\phi_0$ \cite{Herr1, Pascal}. That is to say, all the spectral parameters ($f_c$, $f_{spc}$ and $\phi_m$) of a DKS microcomb can be precisely manipulated by tuning the pump laser frequency (
i.e., the effective pump-cavity detuning $\delta$). Relying on these effects, here we demonstrate coherence-cloned DKS microcomb re-generation by relaying the pump laser between a pair of transmitter and receiver separated by 50 km, with the assistant of a second pilot tone to achieve further mode spacing stabilization and phase noise suppression between the original and re-generated microcombs, based upon the mechanisms of two-point locking and optical frequency division. We illustrate that the re-generated receiver microcomb achieves excellent frequency and phase consistency with the transmitter microcomb, enabling high performance and energy-saving coherent data transmission with substantially simplified processing for the carrier-LO frequency offsets and phase drifts.

\section*{Results}
\textbf{Coherence-cloned re-generation of DKS microcomb.} Our experiments utilize two silicon nitride micro-ring cavities with similar free spectral range (FSR) of $\sim$ 100 GHz \cite{Yong1}. A low-noise fiber laser with wavelength $\lambda_{\rm{pump}}\sim$1550.0 nm is used as the pump laser $C_{\rm{Tx}}(0)$ to produce a DKS microcomb $C_{\rm{Tx}}(m)(m=\pm1, 2, 3,...)$ in the transmitter microcavity, via the technique of auxiliary laser heating (ALH) (see Methods) \cite{Yong2,Heng,Shuangyou}. ALH is adopted in order to suppress the thermal nonlinearity of microcavity resonances and allow the pump laser to stably access single soliton state in the red-detuning regime. Afterwards, the transmitter microcomb $C_{\rm{Tx}}$ together with the pump laser $C_{\rm{Tx}}(0)$ is sent through a 50 km standard-single-mode-fiber (SSMF) to the receiver, where the conveyed pump laser $C_{\rm{Tx}}(0)$ is used to re-generate another DKS microcomb $C_{\rm{Rx}}(m=\pm1, 2, 3,...)$ in the receiver microcavity, also using ALH method \cite{Heng}. Optical spectra of the transmitter microcomb $C_{\rm{Tx}}$ and receiver microcomb $C_{\rm{Rx}}$ are shown in Fig. 1b. Once generated, the microcombs can operate for weeks maintained by stabilization techniques \cite{Yong1}. At this stage, both microcombs have identical central wavelength $\lambda_{\rm{pump}}$, and the comb line phases $\phi_m^{\rm{Rx}}$ and $\phi_m^{\rm{Tx}}, (m=\pm1, 2, 3,...)$ all align approximately to the corresponding pump laser phase $\phi_0^{\rm{Rx}}$ and $\phi_0^{\rm{Rx}}$ respectively \cite{Pascal}. However, due to the distinct soliton repetition rates ($f_{\rm{spc}}^{\rm{Tx}}\sim100.53$ GHz, $f_{\rm{spc}}^{\rm{Rx}}\sim100.58$ GHz) and their uncorrelated jitters caused by the fluctuations of the two independent microcavities, the frequency and phase coherence between the comb lines within $C_{\rm{Tx}}$ and $C_{\rm{Rx}}$ are still weak \cite{Chengying,Tara}, exhibiting inter-comb beat note spectrum with full-width-half-maximum (FWHM) linewidth $>3$ kHz (e.g., for $m=1$), as shown in Fig. 1c. Therefore, at this stage $C_{\rm{Rx}}$ is not yet a coherence-cloned copy of $C_{\rm{Tx}}$. 

Then, we implement phase locking of the 17th receiver comb line $C_{\rm{Rx}}(17)$ to the arrived 17th transmitter comb line $C_{\rm{Tx}}(17)$ (see Methods), and narrow their beat note FWHM linewidth down to $\rm\sim$5 Hz. By doing this, $C_{\rm{Tx}}$ and $C_{\rm{Rx}}$ are two-point locked \cite{Fortier} by the shared pump laser $C_{\rm{Tx}}(0)$, and the 17th comb modes $C_{\rm{Tx}}(17)$ and $C_{\rm{Rx}}(17)$, thus the inter-comb frequency and phase noise of those in-between comb lines $C_{\rm{Tx}}(m)$ and $C_{\rm{Tx}(m)}$ $(m=\pm1, 2, 3,...16)$ can be substantially suppressed obeying the law of optical frequency division (OFD) \cite{Yuan,Qifan,Beichen}. As shown in Fig. 1c, after two-point locking, the linewidths of inter-comb beat notes $(m=1, 5, 10)$ significantly decrease and their noise background varies approximately as the function of $1/(m-17)^2$ \cite{Fortier,Qifan}. Minor close-in ($<$ 3.0 kHz) noises are observed in the beat note spectra due to dispersion incurred temporal walk off among different comb lines 
as $C_{\rm{Tx}}$ transmitted through the 50 km SSFM (see Methods), which, however, have no influence to the performance of coherent data detection, as will be discussed below. Fig. 1d shows the Allan deviation of the beat note frequency $\Delta f_{\rm{CL}}(m)$ between $C_{\rm{Tx}}(m)$ and $C_{\rm{Rx}}(m)$ $(m=1, 5)$, it is seen that two-point locking improves the stability of $\Delta f_{\rm{CL}}$ by about 4 orders of magnitude (at 1 s gate time) comparing with the situation without two-point locking. Of note, here we choose $m=17$ as the locked comb mode due to the bandwidth limitation ($<$ 1.0 GHz) of our phase comparator. According to OFD theory, the locked comb mode index should and can be further increased by adopting either faster phase comparator or smaller difference between $f_{\rm{spc}}^{\rm{Tx}}$ and $f_{\rm{spc}}^{\rm{Rx}}$, so that to obtain larger division factor and stronger coherence enhancement between $C_{\rm{Rx}}$ and $C_{\rm{Tx}}$. Nevertheless, the low noise beat notes and stable Allan deviations shown in Fig. 1c-d indicate that the mutual coherence between the DKS microcombs $C_{\rm{Tx}}$ and $C_{\rm{Rx}}$ is already high. Next, we will show how the highly coherent microcombs facilitate coherent data transmission. 
\newline
\newline
\noindent \textbf{Coherent data interconnect using coherence-cloned microcombs.} Fig. 2a illustrates the experimental setup of optical data interconnect using the coherence-cloned DKS microcombs as carriers and LOs \cite{Yong2}. At the transmitter, 20 comb lines $C_{\rm{Tx}}(m=\pm2...\pm11)$ are selected and sent into an IQ modulator (Fig. 2c), where 21 Gbaud/s single polarization 16-QAM data are encoded to all the comb lines. The modulated data channels together with the pump laser $C_{\rm{Tx}}(0)$ and pilot tone $C_{\rm{Tx}}(17)$ are combined and sent to the receiver through 50 km SSMF. At the receiver, microcomb $C_{\rm{Rx}}$ is re-generated and two-point locked to $C_{\rm{Tx}}$ following the process described above, and then used as LOs for coherent date receiving (Fig. 2c). Of note, as $C_{\rm{Tx}}(0)$ and $C_{\rm{Tx}}(17)$ propagate through the 50 km fiber link together with the high speed data signals, a matter of concern is that the data signals may impose linewidth broadening to them via nonlinear cross-phase modulation (XPM) and degrade their spectral purity as the pump laser and reference pilot. Nevertheless, as shown in Fig. 2d-2e, after co-propagating with data channels, the beat note between $C_{\rm{Tx}}(17)$ and $C_{\rm{Rx}}(17)$ remains almost identical with the case without co-propagating data, indicating that XPM only induces negligible linewidth distortion to $C_{\rm{Tx}}(17)$ and $C_{\rm{Tx}}(0)$ in our experiment. The underlying mechanism is because fiber dispersion induces spatiotemporal walk-off among signals at different wavelengths along the transmission link \cite{lorences, lundberg}, XPM imposed to $C_{\rm{Tx}}(17)$ and $C_{\rm{Tx}}(0)$ from different data channels are smoothed out as a quasi-constant phase envelop without high frequency component (see Supplementary Information for numerical analysis). 

Fig. 2b shows the performance of coherent data receiving using  $C_{\rm{Tx}}$ and $C_{\rm{Rx}}$ as carriers and LOs, it is seen that excellent signal-to-noise ratio (SNR) and bit-error rate (BER) are achieved for all the 20 data channels, summing up to a total line rate of 1.68 Tbit s$^{-1}$. More importantly, thanks to the high coherence between $C_{\rm{Tx}}$ and $C_{\rm{Rx}}$, DSP-based electrical frequency offset estimation (FOE) and carrier phase estimation (CPE) between carriers and LOs are substantially simplified during coherent data retrieval \cite{Ezra,Xiang, Seb,Benjamin}. First, after two-point locking, the frequency offset $\Delta f_{\rm{CL}}(17)$ between $C_{\rm{Rx}}(17)$ and $C_{\rm{Tx}}(17)$ has been locked to the reference clock $f_{\rm{REF}}= 941.101000$ MHz, with tiny residual frequency jitter $<$ 1 Hz (at 100 ms gate time, see Fig. 1d), so the FOE between $C_{\rm{Tx}}(m)$ and $C_{\rm{Rx}}(m)$ $(m=\pm1, 2, ...16)$ can be precalculated using the simple relation $\Delta f_{\rm{CL}}(m)=m\cdot f_{\rm{ref}}/17$. To validate this scheme, we conduct coherent data demodulation using $\Delta f_{\rm{CL}}(m)$ as the FOE result for each channel, and resolve the frequency offset error by extracting the first-order derivative of time from the data phase evolution (see Fig. 2f) \cite{lundberg}. As summarized in Fig. 2g, the discrepancies between the precalculated $\Delta f_{\rm{CL}}(m)$ and actual frequency offsets are within $\pm$500 Hz for all the 20 data channels. In comparison, when $C_{\rm{Rx}}$ and $C_{\rm{Tx}}$ are not two-point locked, the errors of $\Delta f_{\rm{CL}}(m)$ become $\sim$3 orders of magnitude bigger. Hypothetically, if traditional DSP algorithms for FOE (e.g., 4th power spectrum peak search) are used to achieve the accuracy of $\pm$500 Hz, it would entail unacceptably heavy DSP operations and super-long data sequence \cite{Seb}. Thus, by virtue of the mutually locked frequency between $C_{\rm{Tx}}$ and $C_{\rm{Rx}}$, we can save substantial DSP power and complexity while simultaneously achieving FOE accuracy that is orders of magnitude higher than relying on conventional digital methods.   

Second, besides FOE, to retrieve data information from coherently modulated signal, random phase drift rooted in the residual frequency offset and intrinsic linewidths between the carrier and LO needs to be traced using CPE algorithms \cite{Ezra,Seb}. Essentially, the processing rate and corresponding power consumption of CPE depend on the phase coherence between the carrier and LO tones \cite{Benjamin}. In other words, the lowest CPE rate should be properly chosen to minimize power consumption (considering that typical CPE algorithms such as Viterbi$\&$Viterbi phase estimation and blind-phase search are usually power hungry), while making sure that stochastic phase drift within the interval between two CPE operations causes acceptable data bit error \cite{Seb, Benjamin}. It has been shown above that the phase coherence between $C_{\rm{Rx}}$ and $C_{\rm{Tx}}$ is greatly enhanced by two-point locking and OFD, so it is expected that the CPE rate and related power budget can be reduced when using them as carriers and LOs. As shown in Fig. 3b, the data channels carried and demodulated by $C_{\rm{Rx}}$ and $C_{\rm{Tx}}$ exhibit extremely stable phase evolution (see Methods for algorithm used for phase retrieval), with the phase fluctuation much smaller than generated by unlocked microcombs. Larger indexed data channels show slightly bigger phase fluctuations as the corresponding comb lines experienced smaller frequency division factors, but are still confined in a small range (e.g., $<\pm0.2$ rad). Then we gradually slow down the CPE rate (i.e., increasing the number of skipped data blocks after one CPE) while continuously recording the BER, and evaluate the lowest CPE rate allowed for different data channels. As shown in Fig. 3c, if we set the $7\%$ hard forward-error-correction (FEC) threshold 3.8e-3 as the target BER, two-point locked $C_{\rm{Rx}}$ and $C_{\rm{Tx}}$ enable 1 order of magnitude lower CPE rate than unlocked microcombs, and 3 orders of magnitude lower CPE rate than independent carrier and LO lasers. Particularly, for those lower indexed data channels (e.g., $m<5$), only one CPE block (32 symbols) is sufficient to warrant satisfying BER of the whole data frame (400,000 symbols), implying substantial power saving of the relevant DSP module. Practically, such stable phase evolutions between coherence-cloned $C_{\rm{Rx}}$ and $C_{\rm{Tx}}$ can even be tracked by adaptive equalizer module without conducting CPE, and possibly to bring about further simplification to the coherent receiver. Detailed circuits design of coherent receiver that fully copes with coherence-cloned microcombs beyond the scope of the current work, but will be an important topic as Kerr microcomb technology moving fast towards utility.  

Furthermore, Fig. 3e shows the data phase evolutions of different channels that are simultaneously demodulated in two coherent receivers, it is observed that strong phase correlations exist among channels. Such phenomenon can be interpreted by Eq. 10 in Methods section. It shows that fast phase fluctuations between two-point locked microcombs $C_{\rm{Rx}}$ and $C_{\rm{Tx}}$ mainly result from the term $m\cdot\Delta\phi/17$, which linearly scales up with $m$ meaning that we can use the CPE result of one channel to predict the phase of other channels in a master-slave fashion, as sketched in Fig. 3d \cite{Lars, lundberg}. Fig. 3f shows the measured data receiving performance when master-slave CPE is conducted among channel 1 to channel 10, excellent BER is achieved when the phase of higher indexed data channel (e.g., channel 10) is used to detect lower indexed data channels (e.g., channel 1 to 9). For example, when the CPE result of channel 10 is used for channel 1, only minor BER penalty is observed comparing with the result of independent CPE. Moreover, comparing with recent demonstrated master-slave CPE using uncorrelated carrier comb and LO comb \cite{lundberg}, coherence-cloned DKS comb lines possess much longer mutual coherent time, so they should be less sensitive to phase de-coherence caused by fiber dispersion, thus can potentially support longer transmission distance (see Supplementary Information for detailed analysis). 

\section*{Discussion}
Synthesizing the results in Fig. 3c and 3e, we can choose a desired trade-off between BER performance and CPE simplicity according to specific system requirements. For example, if our system has a target BER of 3.8e-3, we can run CPE every other 1001 data block (i.e., skip 1000 blocks after each CPE) for channel 10 and use the CPE result to detect channel 9 to channel 1. So, only 13 CPE operations ($12,500\div1000$) are needed for channel 10 and the total decoded symbol number sums up to 4,000,000. In comparison, if independent carriers and LOs are used,  12,500 CPE operations ($12,500\div10\times 10$ channels) are needed to reach the 3.8e-3 BER (i.e., skipping 10 data blocks after each CPE, see Fig. 3c) within 4,000,000 symbols. According to such evaluation, 10 data channels carried and detected by coherence-cloned microcombs ($i=1000, j=9$) bring about more than 3 orders of magnitude less pilot symbols and related CPE operations comparing with same data capacity implemented by individual laser carriers and LOs ($i=10, j=0$). Such prominent source saving can further scale up when the number of data channel increases.

In summary, we demonstrated coherence-cloned re-generation of DKS microcombs over long distance and used them as the transmitter carriers and receiver LOs for terabit coherent data interconnect. Enabled by two-point locking and OFD, excellent frequency and phase coherence are achieved between the original and re-generated microcombs, which are leveraged to implement totally saving of FOE and substantially reducing of CPE in coherent data detection. In our experiment, point-to-point interconnect is demonstrated, for which the pump laser and pilot tone can be conveyed from transmitter to the receiver. Indeed, such scheme would become difficult for networks with multiple nodes and sophisticated topology \cite{Mikael}. However, instead of conveying pilot tones, two-point locking among transmitter and receiver microcombs can also be implemented using local optical frequency standard such as atomic gas cell or ultra-stable optical cavities \cite{blumenthal}. As long as the mutual stability among the frequency standards at different network nodes is sufficiently high, the above discussed benefits regarding FOE and CPE for coherent data receiving can be obtained, providing a potential solution to cope with the impending energy crisis that vexes the optical transmission industry. 

\vspace{12 pt}
\noindent\textbf{Acknowledgments} The authors acknowledge VLC Photonics S. L. and LiGenTec SA for device fabrication. This work is supported by National Key Research and Development Program of China (2019YFB2203103, 2018YFA0307400); National Natural Science Foundation of China (61705033, 61775025); The 111 project (B14039).

\section*{Methods}
\textbf{Generation and locking of $C_{\rm{Tx}}$ and $C_{\rm{Rx}}$.} DKS microcomb $C_{\rm{Tx}}$ is first generated in the transmitter microcavity, using the auxiliary laser heating method \cite{Heng, Shuangyou}. Particularly, an auxiliary laser is tuned into the blue-detuning regime of a cavity mode ($\sim1536.2$ nm), and subsequently a pump laser is tuned into another cavity mode ($\sim1549.9$ nm). By properly setting the power and detuning of the pump and auxiliary laser, the heat flow caused by them can be balanced out allowing the pump laser to stably scan into the red-detuning regime and access single soliton state \cite{Heng}. Moreover, using the same pump laser sent from the transmitter to the receiver trough 50 km SSMF, DKS microcomb $C_{\rm{Rx}}$ is similarly generated in the receiver microcavity, by using another auxiliary laser ($\sim1536.2$ nm) to simultaneously control the pump detuning and maintain cavity thermal stability \cite{Yong1}.  

To achieve two-point locking, $C_{\rm{Tx}}(17)$ and $C_{\rm{Rx}}(17)$ are filtered out and sent into a fast photodiode in which their beating frequency $\Delta f_{\rm{CL}}(17)$ is detected. Then, $\Delta f_{\rm{CL}}(17)$ and a reference clock $ f_{\rm{REF}}=941.101000$ MHz is sent into a phase comparator to generate the error signal and feedback control the power of the auxiliary laser for generating $C_{\rm{Rx}}$. Particularly, the auxiliary laser power controls the pump detuning in the receiver microcavity via thermal resonance shift and in turn adjust the repetition rate of $C_{\rm{Rx}}$ \cite{Chengying,Yong1}, so that to lock $C_{\rm{Rx}}(17)$ to $C_{\rm{Tx}}(17)$. The bandwidth of our phase lock loop is about 100 kHz, set by the amplitude modulation frequency limitation of the adopted auxiliary laser module. 

\noindent \textbf{Coherence analysis between $C_{\rm{Tx}}$ and $C_{\rm{Rx}}$.} 
After 50 km fiber transmission, the phase of the pump laser when it arrives at the receiver side is\cite{lorences}: 
\begin{center}
\begin{equation}
\phi^{\rm{Tx(0)}}=\phi_{\rm{int}}^{\rm{Tx(0)}}+\phi_{\rm{ff}}^{\rm{Tx(0)}}+\phi_{\rm{nl}}^{\rm{Tx(0)}} 
\end{equation}
\end{center}

\noindent $\phi_{\rm{int}}^{\rm{Tx(0)}}$ is the intrinsic phase noise (i.e., linewidth) of the pump laser, $\phi_{\rm{ff}}^{\rm{Tx(0)}}$ denotes the phase noise caused by the random fluctuation of 50 km SSMF, $\phi_{\rm{nl}}^{\rm{Tx(0)}}$ denotes the nonlinear phase modulation acquired during fiber transmission. Note that all the phase terms in Eq. 1 are time-varying. Similarly, the phase of the $m$th transmitter comb line $C_{\rm{Tx}}(m)$ when it arrives at the receiver side is:
\begin{center}
\begin{equation}
\phi^{\rm{Tx(\textit{m})}}=\phi_{\rm{int}}^{\rm{Tx(\textit{m})}}+\phi_{\rm{ff}}^{\rm{Tx(\textit{m})}}+\phi_{\rm{nl}}^{\rm{Tx(\textit{m})}}+\phi_{\rm{rep}}^{\rm{Tx(\textit{m})}}
\end{equation}
\end{center}

\noindent Here $\phi_{\rm{rep}}^{\rm{Tx(\textit{m})}} = 2\pi\cdot m\cdot\Delta f_{\rm{rep}}^{\rm{Tx}}\cdot t$ is the phase noise caused by the fluctuation of soliton repetition rate $\Delta f_{\rm{rep}}^{\rm{Tx}}$ of $C_{\rm{Tx}}$. 

At the receiver site, the phase of the $m$th line of the re-generated DKS microcomb $C_{\rm{Rx}}$ is:
\begin{center}
\begin{equation}
\phi^{\rm{Rx(\textit{m})}}=\phi_{\rm{int}}^{\rm{Rx(\textit{m})}}+\phi_{\rm{rep}}^{\rm{Rx(\textit{m})}}
\end{equation}
\end{center}

\noindent Here $\phi_{\rm{rep}}^{\rm{Rx(\textit{m})}} = 2\pi\cdot m\cdot\Delta f_{\rm{rep}}^{\rm{Rx}}\cdot t$, with $\Delta f_{\rm{rep}}^{\rm{Rx}}$ the soliton repetition rate jitter of $C_{\rm{Rx}}$. For $C_{\rm{Rx}}$ we neglect the random length fluctuation and nonlinear phase modulation in those short fiber patch cord within the receiver. 

As $C_{\rm{Rx}}$ and $C_{\rm{Tx}}$ are both in the state of DKS mode locking, we can assume that the intrinsic phase of each comb line is aligned to the phase of corresponding pump laser:

\begin{center}
\begin{equation}
\phi_{\rm{int}}^{\rm{Tx(\textit{m})}}=\phi_{\rm{int}}^{\rm{Tx(0)}}, \phi_{\rm{int}}^{\rm{Rx(\textit{m})}}=\phi^{\rm{Tx(0)}} 
\end{equation}
\end{center}

\noindent Also, the phase variation caused by soliton repetition rate change for $C_{\rm{Rx}}(m)$ and $C_{\rm{Tx}}(m)$ has the relationship: 

\begin{center}
\begin{equation}
\phi_{\rm{rep}}^{\rm{Tx(\textit{m})}} = m\times\phi_{\rm{rep}}^{\rm{Tx(1)}}, \phi_{\rm{rep}}^{\rm{Rx(\textit{m})}} = m\times\phi_{\rm{rep}}^{\rm{Rx(1)}}
\end{equation}
\end{center}

\noindent Locking $C_{\rm{Rx}}(17)$ to $C_{\rm{Tx}}(17)$ leads to: $\phi^{\rm{Tx(17)}}-\phi^{\rm{Rx(17)}}=\Delta\phi$, $\Delta\phi$ is the residual technical noise of the adopted phase lock loop, and we obtain: \\
\begin{center}
\begin{equation}
\phi_{\rm{int}}^{\rm{Tx(17)}}+\phi_{\rm{ff}}^{\rm{Tx(17)}}+\phi_{\rm{nl}}^{\rm{Tx(17)}}+\phi_{\rm{rep}}^{\rm{Tx(17)}}=\phi_{\rm{int}}^{\rm{Rx(17)}}+\phi_{\rm{rep}}^{\rm{Rx(17)}}+\Delta\phi
\end{equation}
\end{center}

\noindent Using Eq(1-6), we can calculate the phase of the $m$th inter-comb beat note:
\begin{center}
\begin{equation}
\phi^{\rm{Tx(\textit{m})}}-\phi^{\rm{Rx(\textit{m})}}= \Delta\phi_{\rm{p}}^{\rm{(m)}}-\Delta\phi_{\rm{l}}^{\rm{(m)}}+m\cdot\Delta\phi/17
\end{equation}
\end{center}

\begin{center}
\begin{equation}
\Delta\phi_{\rm{p}}^{\rm{(m)}}= (\phi_{\rm{ff}}^{\rm{Tx(\textit{m})}}+\phi_{\rm{nl}}^{\rm{Tx(\textit{m})}}-\phi_{\rm{ff}}^{\rm{Tx(0)}}-\phi_{\rm{nl}}^{\rm{Tx(0)}})
\end{equation}
\end{center}

\begin{center}
\begin{equation}
\Delta\phi_{\rm{l}}^{\rm{(m)}}= m\cdot(\phi_{\rm{ff}}^{\rm{Tx(17)}}+\phi_{\rm{nl}}^{\rm{Tx(17)}}-\phi_{\rm{ff}}^{\rm{Tx(0)}}-\phi_{\rm{nl}}^{\rm{Tx(0)}})/17
\end{equation}
\end{center}

\noindent $\Delta\phi_{\rm{p}}^{\rm{(m)}}$ and $\Delta\phi_{\rm{l}}^{\rm{(m)}}$ are produced due to that, when $C_{\rm{Rx}}$ is generated the pump laser has been attached with extra phase noise $\phi_{\rm{ff}}^{\rm{Tx(0)}}+\phi_{\rm{nl}}^{\rm{Tx(0)}}$ comparing with the original pump laser linewidth $\phi_{\rm{int}}^{\rm{Tx(0)}}$ when $C_{\rm{Tx}}$ is generated. Since in our experiment the nonlinear XPM among different channels are small within the 50 km SSFM (see Supplementary Information), we can neglect the nonlinear phase modulation term $\phi_{\rm{nl}}^{\rm{Tx(\textit{m})}}$ $m=\pm1,2,3,...$ and obtain:

\begin{center}
\begin{equation}
\phi^{\rm{Tx(\textit{m})}}-\phi^{\rm{Rx(\textit{m})}}= (\phi_{\rm{ff}}^{\rm{Tx(\textit{m})}}-\phi_{\rm{ff}}^{\rm{Tx(0)}})-m\cdot(\phi_{\rm{ff}}^{\rm{Tx(17)}}-\phi_{\rm{ff}}^{\rm{Tx(0)}})/17+m\cdot\Delta\phi/17
\end{equation}
\end{center}

\noindent In the presence of fiber dispersion, different comb modes temporally walk off among each other, so the instantaneous phases of different comb lines at the output of the SSFM are different: $\phi_{\rm{ff}}^{\rm{Tx(\textit{m})}}\ne\phi_{\rm{ff}}^{\rm{Tx(\textit{n})}}, m\neq n$. This results in the non-zero first and second term to the right-hand-side of Eq. 10. Considering that fiber fluctuations are low frequency, we speculate that the first and second term in Eq. 10 explain the minor low-frequency noise  ($<$ 3.0 kHz) observed in the beat note spectra shown in Fig. 1c. However, in our coherent detection experiment the electrical CPE rate is generally larger than 3.0 kHz, so such slow phase drifts caused by fiber fluctuations hardly influence CPE and data receiving. The last term in Eq. 10 is linearly depends on comb mode index $m$, based on which master-slave CPE is conducted. 

\noindent \textbf{Coherent data modulation and receiving using $C_{\rm{Tx}}$ and $C_{\rm{Rx}}$.} At the transmitter, 20 DKS comb lines are filtered out by a C-band programmable wavelength selective switch (WSS) and used as data  carriers. Each of the 20 comb lines is boosted to about 0 dBm using a low-noise Er-doped fiber amplifier (EDFA) while maintaining $>$40 dB optical carrier-to-noise-ratio (OCNR)(Fig. 2c). An IQ modulator is used to encode single-polarization 16-QAM data onto all the 20 comb lines, driven by an electrical arbitrary waveform generator (eAWG) to generate the 16-QAM waveform with rectangle pulse shaping. After modulation, all data channels are amplified by another EDFA to generate -10.0 dBm launched power for each channel.

The 20 data channels together with the pump laser $C_{\rm{Tx}}(0)$ and pilot tone $C_{\rm{Tx}}(17)$ are transmitted through 50 km SSMF to the receiver. At the receiver, $C_{\rm{Rx}}$ is re-generated as LOs for coherent data detection. Each data channel and the corresponding LO is selected by another WSS and fed into a coherent receiver. The detected electrical signal of each channel is recorded by a real-time digital phosphor oscilloscope (DPO) and then processed offline. Multiple algorithms are used to achieve optimal data retrieval, including IQ imbalance compensation based on Gram-Schmidt orthogonalization, chromatic dispersion compensation, Volterra channel equalization, and pilot-aided CPE. Particularly, CPE is implemented by comparing the phase of 1 data block (32 symbols) of the received data sequence with the originally sent pilot symbols. For CPE investigation presented in Fig. 3, two coherent receivers and four DPO channels are used to simultaneously receive two 16-QAM data channels modulated at 12.5 Gband (i.e., 400,000 symbols or 12,500 data blocks within the 32 $\rm{\mu s}$ time window), so as to reserve their phase coherence for master-slave joint processing.

\end{spacing} 

\normalem
\bibliographystyle{unsrt}
\bibliography{references}


\newpage

\begin{figure}[ht]
\centering
\includegraphics[scale=0.33]{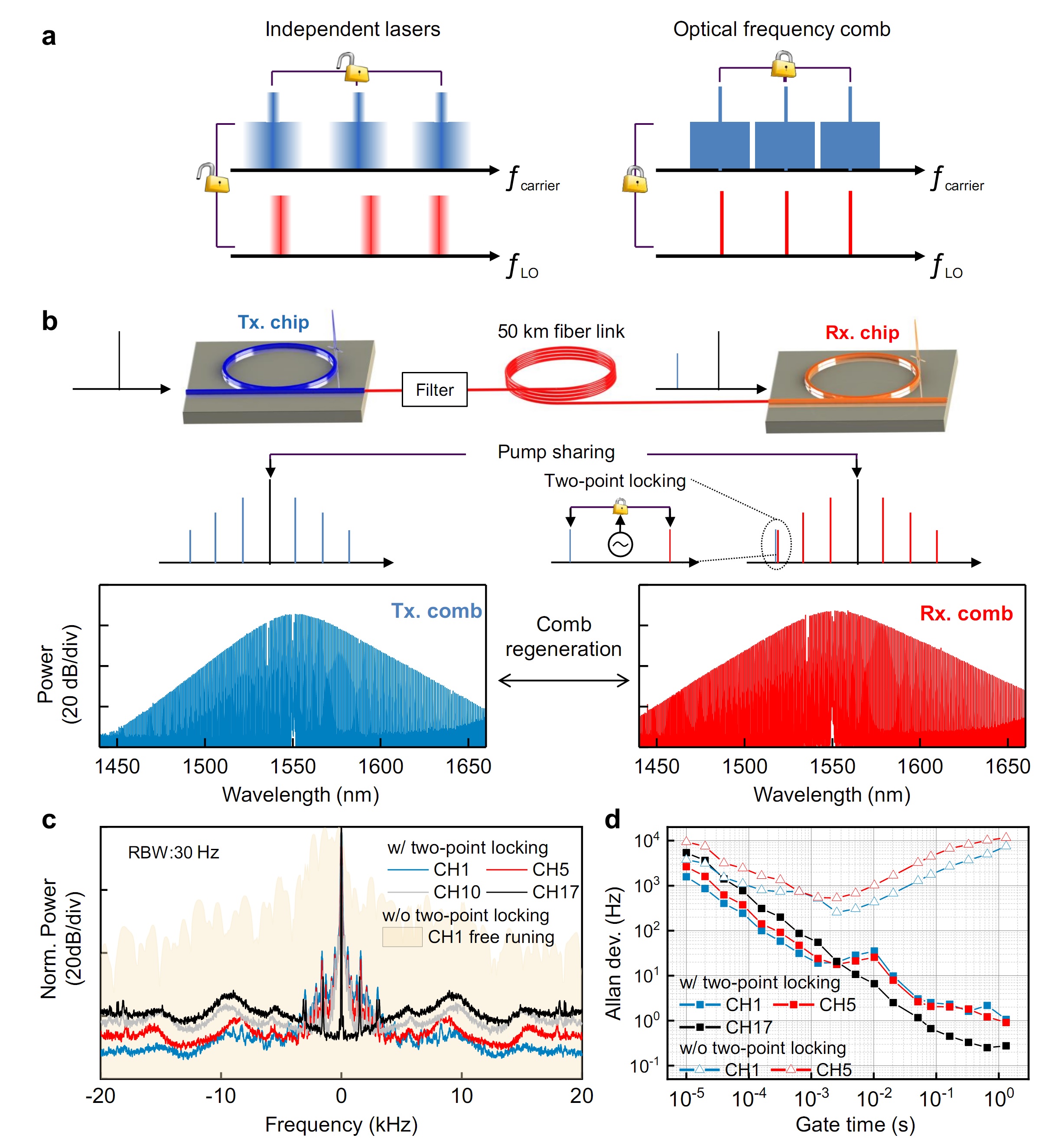}
\caption{\textbf{Coherence-cloned re-generation of DKS microcomb.} \textbf{a} Left: for conventional WDM system based on independent lasers, guard intervals are necessary to tolerate the random frequency drifts among adjacent channels at the transmitter side, while at the receiver side power-hungry DSP must be implemented to recovery and compensate the random frequency and phase drifts between the carriers and LOs. Right: optical frequency comb has much better spectral stability than independent lasers, thus holds great potentials to improve spectral efficiency by eliminating guard intervals and reduce the DSP complexity of coherent transmission. \textbf{b} Upper: schematics of coherence-cloned DKS microcomb re-generation and two-point locking. Lower: optical spectra of the original transmitter comb $C_{\rm{Tx}}$ and re-generated receiver comb $C_{\rm{Rx}}$. \textbf{c} Inter-comb beat note spectra between $C_{\rm{Tx}}(m)$ and $C_{\rm{Rx}}(m)$, $m=1, 5, 10, 17$. It is seen that after two-point locking by the conveyed pump laser and pilot-tone, the beat note linewidths are substantially narrowed down, implying that coherence between $C_{\rm{Tx}}$ and $C_{\rm{Rx}}$ are significantly enhanced comparing with the situation without two-point locking. \textbf{d} Allan deviation of the inter-comb beat note frequency $\Delta f_{\rm{CL}}(m)$, $m=1, 5, 17$, confirming the efficacy of frequency stability enhancement by two-point locking.  The fundamental repetition rate offset $\Delta f_{\rm{CL}}(1)=55.358882$ MHz.}
\label{fig:Figure1}
\end{figure}

\newpage

\begin{figure}[ht]
\centering
\includegraphics[scale=0.3]{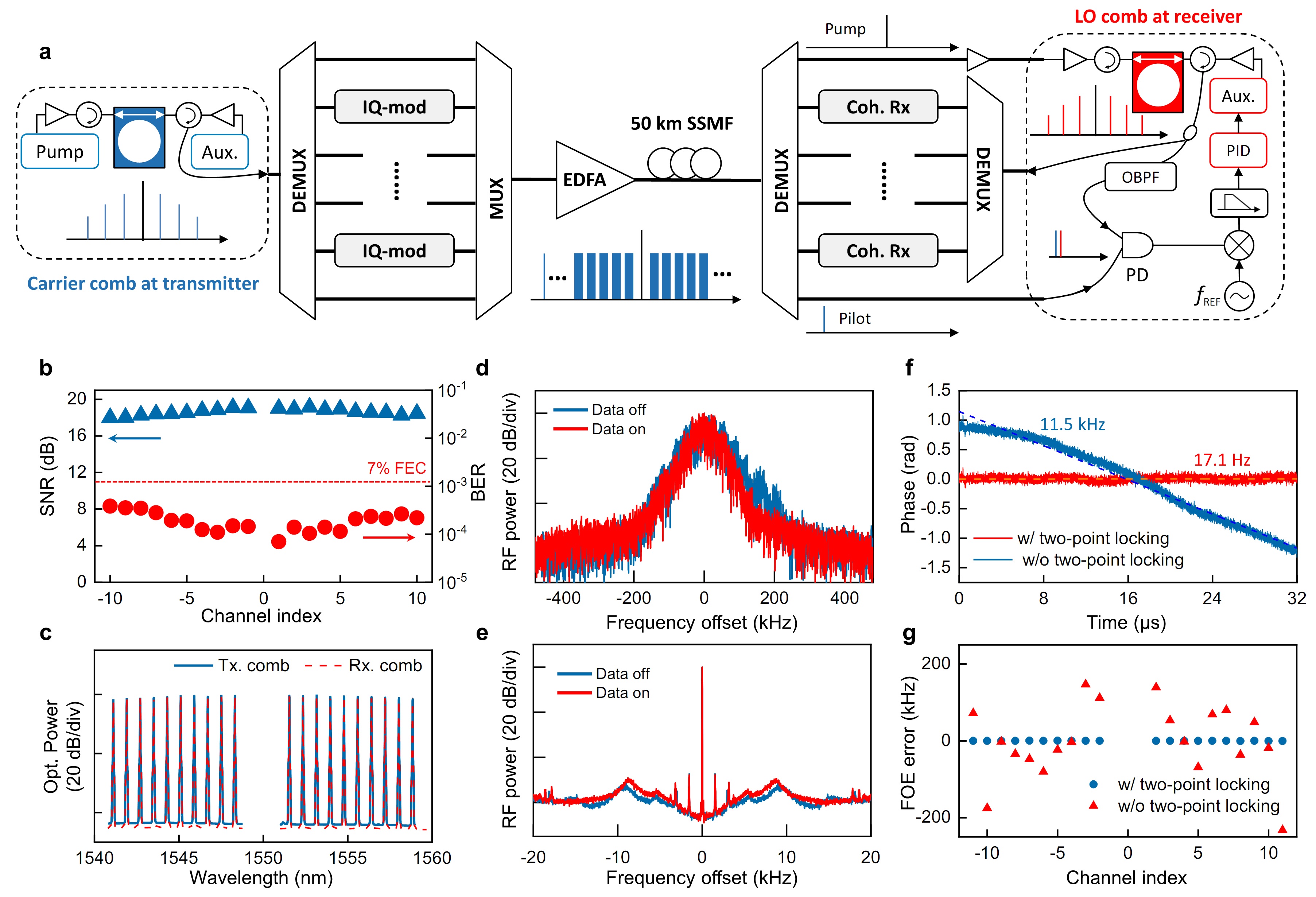}
\caption{\textbf{Optical interconnect using coherence-cloned DKS microcombs as carriers and LOs.} \textbf{a} Setup of the optical coherent data interconnect experiment. At the transmitter, microcomb $C_{\rm{Tx}}$ is generated in a silicon nitride micro-ring cavity based on the ALH method and used as multi-wavelength laser carriers. A programmable de-multiplexer (DEMUX) is adopted to select 20 comb lines from $C_{\rm{Tx}}$ and sent them to a high speed IQ modulator, where 21 Gbaud 16-QAM data is encoded to all the comb lines, forming 20 data channels with total bitrate 1.68 Tbit s$^{-1}$. The pump laser $C_{\rm{Tx}}(0)$, pilot-tone $C_{\rm{Tx}}(17)$ and 20 data channels are then combined together in a multiplexer (MUX) and sent to the receiver through 50 km SSMF. At the receiver, microcomb $C_{\rm{Rx}}$ is re-generated and two-point locked to $C_{\rm{Tx}}$ and used as LOs for coherent data receiving. PID: proportional integral derivative; PD: photodiode; OBPF: optical bandpass filter. \textbf{b} SNR and BER measurements of 50 km interconnect experiment for all the 20 data channels. \textbf{c}, Optical spectra of the carrier and LO comb lines, showing high OCNR $>40$ dB for each line. \textbf{d-e} Comparison of beat note spectra between $C_{\rm{Tx}}(17)$ and $C_{\rm{Rx}}(17)$ before (\textbf{d}) and after (\textbf{e}) two-point locking. The results indicate that nonlinear XPM from the high speed data imposes negligible linewidth distortion to the pilot-tone $C_{\rm{Tx}}(17)$ comparing with the case without data signal, therefore does not impact the performance of microcomb re-generation and coherent data receiving (see supplementary for more discussions). \textbf{f} The resolve phase evolution of channel 1 when FOE is directly calculated via $m\cdot f_{\rm{REF}}/17$. The dotted lines show the residual FOE error extracted from the first-order derivative of time from the data phase evolution, showing a 17.1 Hz (11.5 kHz) FOE error with two-point locking (without two-point locking). \textbf{g} Summarized FOE errors with and without two-point locking for all the 20 data channels.}
\label{fig:Figure2}
\end{figure}

\begin{figure}[ht]
\centering
\includegraphics[scale=0.33]{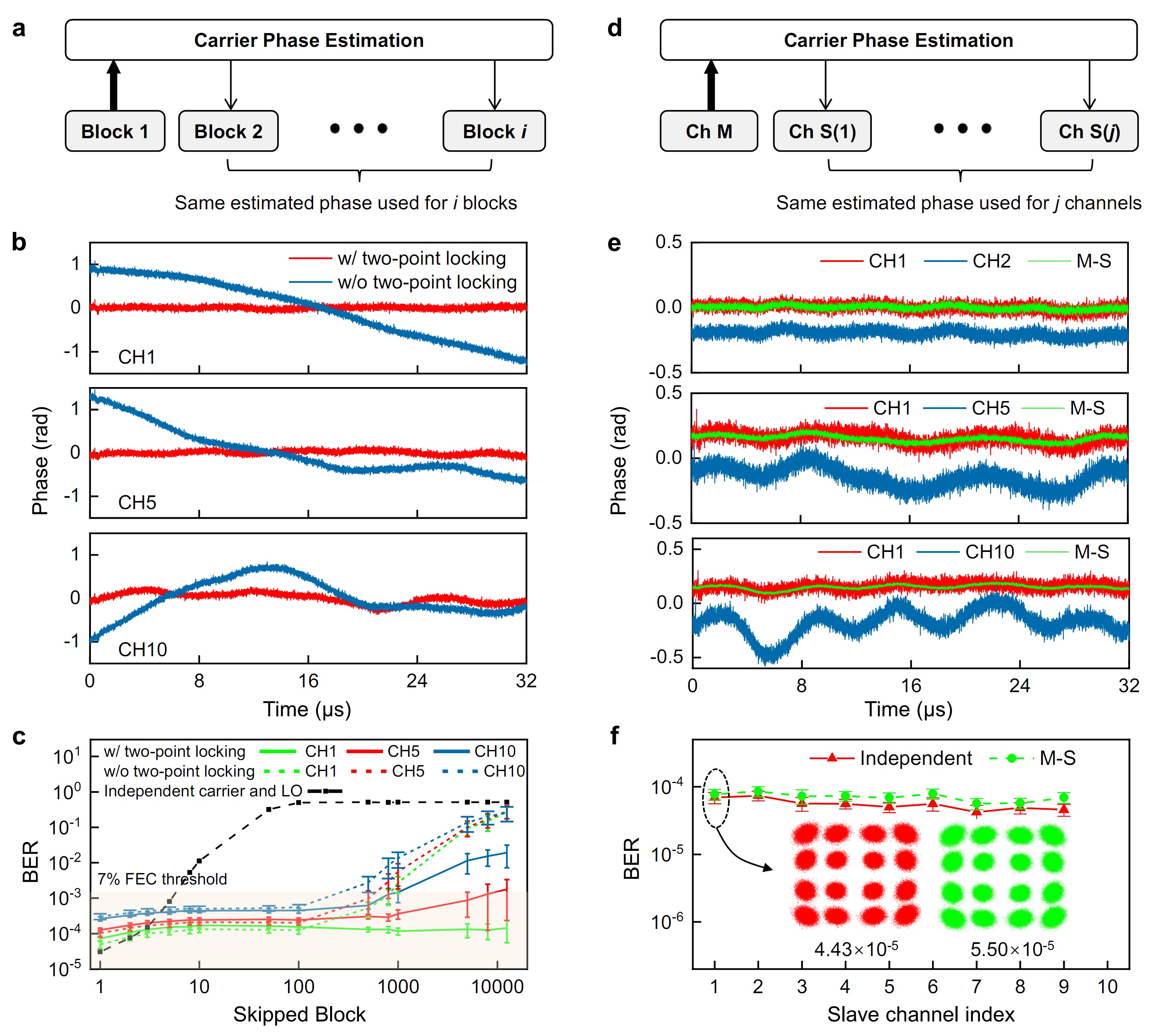}
\newpage
\caption{\textbf{Carrier phase estimation facilitated by coherence-cloned DKS microcombs.} \textbf{a} Scheme for CPE rate configuration. CPE is conducted once every $(i+1)$ data blocks, namely, the CPE results of block 1 is used for the following $i$ data blocks. \textbf{b} CPE results for channel 1,  5 and 10. It is seen that two-point locking between $C_{\rm{Tx}}$ and $C_{\rm{Rx}}$ significantly enhance the phase stability between carriers and LOs. For this measurement the data format is 12.5 Gbaud 16-QAM, each panel illustrates a time window of 32 us containing 400,000 symbols. \textbf{c} Measured BER as a function of CPE rate. The block size for pilot-based CPE is 32 symbols. It is obvious that coherence-cloned microcombs allow much slower CPE rate to reach the target BER 3.8e-3 comparing with unlocked microcombs and independent carriers and LOs. \textbf{d} Scheme for master-slave joint CPE among multiple data channels. The carrier phase is estimated from the master channel (CH M) and then applied to $j$ slave channels (CH S). \textbf{e} Retrieved data phases by individual CPE and master-slave CPE based on Eq. 10. Upper: channel 1 as slave and channel 2 as master; Middle: channel 1 as slave and channel 5 as master; Lower: channel 1 as slave and channel 10 as master. \textbf{f} Summarized BER performance of individual CPE and master-slave CPE of multiple data channels. For master-slave CPE measurement, channel 10 is used as the master channel and channel 1 to 9 are processed as slave channels. The inset shows the constellation maps and BER for channel 1 retrieved via individual and master-salve CPE.}
\label{fig:Figure3}
\end{figure}

\end{document}